\PassOptionsToPackage{hyphens}{url}
\documentclass[aps,prl,reprint,superscriptaddress,nofootinbib]{revtex4-2}
\usepackage[T1]{fontenc}
\usepackage[utf8]{inputenc}
\usepackage{amsmath,amssymb,bm}
\usepackage{graphicx}
\usepackage{booktabs}
\usepackage{array}
\usepackage{xcolor}
\usepackage{url}
\newcommand{\avg}[1]{\langle #1 \rangle}

\newcommand{\XiEff}{\Xi_{\rm eff}}
\newcommand{\XiZero}{\Xi_0}
\newcommand{\Bin}{B_{\rm in}}
\newcommand{\Dz}{D_z}
\newcommand{\Dx}{D_x}
\newcommand{\Tr}{T_{\rm r}}
\newcommand{\Peff}{P_{\rm eff}}
\newcommand{\Peffc}{P_{{\rm eff},c}}
\newcommand{\rL}{r_{\rm L}}

\newcommand{\btr}{\beta_{\rm tr}}
\newcommand{\de}{d_e}
\newcommand{\tausp}{\tau_{\rm sp}}

\newcommand{\Jm}{\mathbf J_M}
\newcommand{\safeincludegraphics}[2][]{%
  \IfFileExists{#2}{\includegraphics[#1]{#2}}{\fbox{\parbox{0.86\linewidth}{\centering Missing figure file: \texttt{\detokenize{#2}}}}}%
}

\begin{document}

\title{Stern--Gerlach Spin Sorting and Dynamical Feedback in Relativistic Pair-Plasma Reconnection}

\author{K. Nykyri}
\affiliation{Physical Sciences Department and Centre for Space and Atmospheric Research, Embry--Riddle Aeronautical University, Daytona Beach, Florida 32114, USA}

\begin{abstract}
We derive a Stern--Gerlach control parameter, $\Xi$, comparing
spin-driven cross-sheet displacement with the relativistic Larmor
radius. It places heliospheric plasmas and most astrophysical jets in the negligible regime,
some stellar-mass black-hole coronae in a transitional regime, and
magnetar sheets in a strong, near-QED regime. Relativistic pair-plasma
simulations at fixed $\gamma_{\rm tr}=2$ show that increasing $\Xi$
produces, within the coupled SG+$\mathbf J_M$ model, magnetic-moment sorting and magnetization-current feedback that enhance the normalized flux-growth rate through an additional spin-kinetic pathway beyond classical pressure- and geometry-controlled pair-plasma reconnection.
\end{abstract}
\maketitle

\begin{figure}[t]
\centering
\safeincludegraphics[width=\columnwidth]{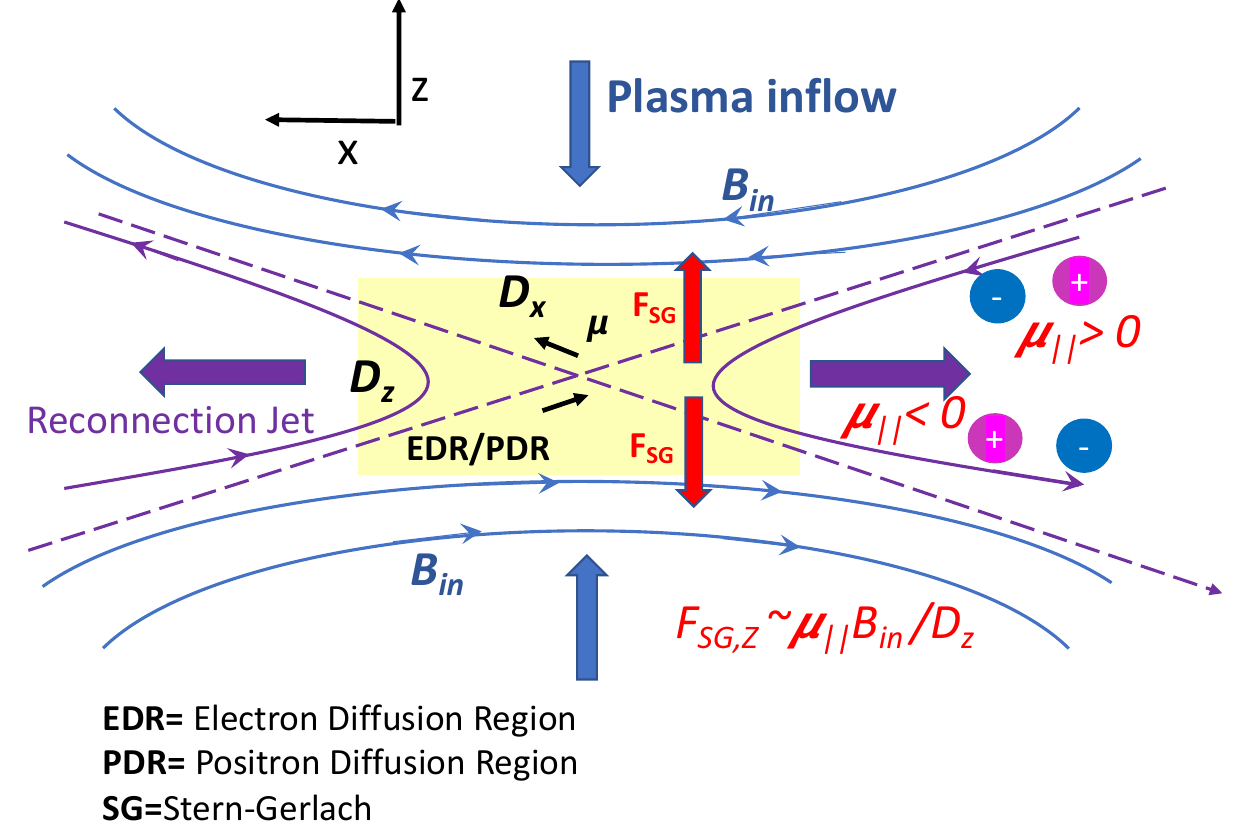}
\caption{Stern--Gerlach quantum spin sorting in the EDR/PDR of a reconnecting current sheet.  The sign convention is \(B_x\simeq+\Bin z/\Dz\), so \(\partial B_x/\partial z>0\); \(+z\) is upward.  Particles with \(\mu_\parallel>0\) and \(\mu_\parallel<0\) are deflected into opposite sides of the current sheet by \(F_{{\rm SG},z}\simeq\mu_\parallel\Bin/\Dz\).  The labels \(+\) and \(-\) denote positrons and electrons, while the red labels denote the sign of \(\mu_\parallel\).  In pair plasmas, particles sharing the same magnetic-moment projection sort into the same \(z\) half-plane even when their spin projections are opposite.}
\label{fig:schematic}
\end{figure}

Magnetic reconnection converts magnetic energy into bulk flow, heating, and nonthermal particle acceleration through a localized nonideal region in which field lines change connectivity~\cite{Shay1999,Drake2014,Dahlin2014,Burch2016} and in diamagnetic cavities, formed in the reconnection outflow regions near cusp-like magnetic field geometries~\cite{Nykyri2019}.  In relativistic plasmas---especially pair plasmas relevant to compact-object coronae, magnetar magnetospheres, and some high-energy laboratory settings---one can ask whether quantum spin physics ever becomes dynamically relevant in the diffusion region~\cite{Sironi2014,YuanNarayan2014,Uzdensky2011}.  Spin-fluid and spin-kinetic plasma theories have existed for years~\cite{Marklund2007PRL,BrodinMarklund2007NJP,Zamanian2010NJP}, but reconnection still lacks a compact regime parameter answering the practical question: when can the Stern--Gerlach (SG) force \cite{GerlachStern1922} compete with classical orbit dynamics?

In this Letter we show that the SG deflection in a reconnecting current sheet can be quantified by a dimensionless ratio.  Second, we show that the same force produces a topology-locked, magnetic-moment-tagged north--south anisotropy in pair plasmas.  We also show that the SG force strongly modifies reconnection dynamics in a Harris-sheet configuration~\cite{Harris1962}, addressed here using SpinPIC2D, a 2.5-dimensional pair-plasma particle-in-cell (PIC) code that advances electrons and positrons in a Harris current sheet with BMT spin precession, a full-step SG momentum kick, a Boris push, and a Yee field update~\cite{YeeGrid1966}. The production implementation combines the SG forcing with the self-consistent magnetization-current closure.  The details of the \(\XiZero\) derivation and the numerical code and its validation are given in the Supplemental Material and in the companion PRE paper~\cite{SM,NykyriPRECompanion}.

The established reconnection rate theories connect fast reconnection to \(x\)-line pressure depletion and the open exhaust geometry it permits.  In nonrelativistic electron--ion plasmas, Liu \emph{et al.} showed that Hall-field-mediated energy transport limits pressure buildup at the \(x\) line, whereas Goodbred and Liu showed that in magnetically dominated relativistic pair plasmas the energy required to sustain the extreme current density can itself deplete the \(x\)-line pressure and drive field collapse~\cite{LiuCassak2022Rate,GoodbredLiu2022}; the resulting normalized rate is constrained by the separatrix geometry~\cite{Liu2017}.  These classical mechanisms are organized by plasma magnetization (not to be confused with the quantum spin magnetization, discussed in this Letter), pressure balance, and geometry rather than by a spin-sensitive field-and-gradient parameter.  The SG channel adds such a parameter: because \(\XiZero\propto\Bin^{2}\) at fixed geometry, spin-kinetic sorting activates in strong-field, strong-gradient sheets.  With the direct \(y\)-component of the SG force vanishing in the 2.5-D geometry (\(\partial/\partial y=0\)), the feedback is indirect: the sheet-normal force reshapes velocity space and pressure moments while the spin-structured populations generate a magnetization current \(J_{M,y}\) that contributes to the reconnecting current system.

For a particle with magnetic moment \(\bm\mu\), the SG force is
\begin{equation}
\mathbf F_{\rm SG}=\nabla(\bm\mu\cdot\mathbf B),
\qquad
\bm\mu=g\frac{q}{2m_e}\mathbf S,
\label{eq:FSG}
\end{equation}
where \(q=-e\) for electrons and \(q=+e\) for positrons.  In the diffusion-region geometry the inflow field is along \(\hat x\), and the dominant current-sheet gradient is \(\partial B_x/\partial z\sim\Bin/\Dz\).  Rapid Bargmann--Michel--Telegdi precession about the inflow field averages away transverse spin projections over a transit~\cite{GriffithsQM,BMT1959}.  The branch-resolved cross-sheet force is therefore
\begin{equation}
F_{{\rm SG},z}\simeq \mu_\parallel\frac{\Bin}{\Dz}
= \operatorname{sgn}(\mu_\parallel)\eta_\mu\mu_B\frac{\Bin}{\Dz},
\label{eq:Fbranch}
\end{equation}
where \(\mu_\parallel\equiv\bm\mu\cdot\hat{\mathbf B}_{\rm in}\), \(\eta_\mu\equiv |\mu_\parallel|/\mu_B\leq1\), and \(\mu_B=e\hbar/(2m_e)\).  The effective ensemble force used for regime estimates is obtained by replacing \(\eta_\mu\) by an effective branch participation/projection factor \(\Peff\):
\begin{equation}
\avg{|F_{{\rm SG},z}|}\simeq \Peff\mu_B\frac{\Bin}{\Dz}.
\label{eq:Favg}
\end{equation}
This notation separates branch sorting, which can occur even for zero net spin polarization, from net macroscopic observables, which require nonzero branch weighting or incomplete cancellation.

For a particle traversing a diffusion region of half-length \(\Dx/2\) at characteristic transit speed \(v_{\rm tr}=\btr c\), the transit time is \(\Tr=\Dx/(2\btr c)\).  The branch displacement for a fully projected moment \((\eta_\mu=1)\) is
\begin{equation}
\avg{z}_{|\mu_\parallel|=\mu_B}=\frac{1}{2}a_z\Tr^2
=\frac{\mu_B\Bin\Dx^2}{8\avg{\gamma}m_e\Dz\btr^2c^2}.
\label{eq:z0}
\end{equation}
Normalizing by the relativistic Larmor radius
\begin{equation}
\rL=\frac{\avg{\gamma}m_e\btr c}{e\Bin}
\end{equation}
gives the branch control parameter
\begin{equation}
\XiZero\equiv\frac{\avg{z}_{|\mu_\parallel|=\mu_B}}{\rL}
=\frac{1}{8}\frac{\mu_Be}{m_e^2c^3}
\frac{\Bin^2\Dx^2}{\avg{\gamma}^{2}\Dz}\,\btr^{-3}.
\label{eq:Xi0}
\end{equation}
For an effective participating population,
\begin{equation}
\XiEff=\Peff\XiZero,
\qquad
\Peffc=\XiZero^{-1}
=8\frac{m_e^2c^3}{\mu_Be}
\frac{\avg{\gamma}^{2}\Dz}{\Bin^2\Dx^2}\btr^3,
\label{eq:Pcrit}
\end{equation}
where \(\Peffc\) is the effective branch factor required for \(\XiEff=1\).  Thus \(\XiEff\ll1\) means SG transport is negligible compared with classical gyration; \(\XiEff\sim1\) marks a transitional regime; and \(\XiEff\gg1\) identifies a strong-field regime in which SG transport cannot be ignored.

The same derivation gives a geometric result.  The SG force sorts particles according to \(\operatorname{sgn}\mu_\parallel\).  With \(\partial B_x/\partial z>0\), \(\mu_\parallel>0\) is driven toward \(z>0\) and \(\mu_\parallel<0\) toward \(z<0\) (Fig.~\ref{fig:schematic}).  In pair plasmas the sorting is most naturally organized by \(\mu_\parallel\): electrons and positrons with the same magnetic-moment projection are deflected into the same half-plane, whereas equal spin projections correspond to opposite \(\mu_\parallel\) and therefore opposite SG deflections.  Figure~\ref{fig:schematic} should therefore be read as moment-tagged, not species-tagged.

Representative evaluations of Eqs.~(\ref{eq:Xi0}) and~(\ref{eq:Pcrit}) are summarized in Table~\ref{tab:survey} of the End Matter.  They place the magnetotail, solar corona, AGN/blazar jets, and Crab pulsar-wind nebula deep in the negligible regime, identify some stellar black-hole-corona geometries as transitional, and show that magnetar current sheets are the natural strong-field targets.  The geometric assumptions and transit-speed closures used in the survey are documented in the Supplemental Material~\cite{SM}.

\begin{figure*}[t]
\centering
\safeincludegraphics[width=0.95\textwidth]{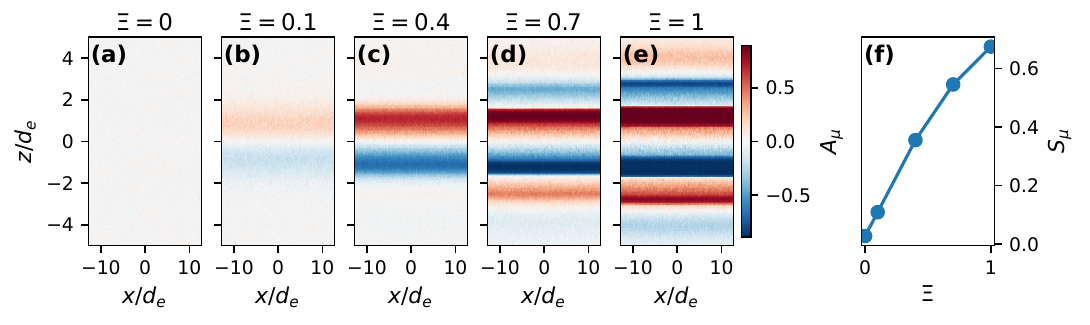}
\caption{\textbf{Controlled onset of Stern--Gerlach branch sorting across the \(\Xi\) scan.}
(a)--(e) Magnetic-moment branch asymmetry \(A_\mu=(n_{\mu+}-n_{\mu-})/(n_{\mu+}+n_{\mu-})\) at \(t\simeq7\tausp\) for the common-\(\gamma_{\rm tr}=2\) scan with \(\Xi=0,0.1,0.4,0.7,1\), shown with one common symmetric color scale; limited saturation is allowed to preserve direct cross-panel comparison of sheet-scale structure.  (f) Density-weighted sorting amplitude \(S_\mu=\sum|n_{\mu+}-n_{\mu-}|/\sum(n_{\mu+}+n_{\mu-})\) in the same central current-sheet region for every run.  The control and \(\Xi=0.1\) cases remain weakly organized, coherent sheet-aligned sorting appears near \(\Xi=0.4\), and strong branch layering develops for \(\Xi=0.7\)--1.}
\label{fig:xi_sorting}
\end{figure*}

To test whether SG sorting can feed back on the reconnecting current sheet, we performed a weak-seed scan with the relativistic SpinPIC2D model at the common transit/coupling anchor \(\gamma_{\rm tr}=2\), varying \(\Xi\) while holding this relativistic reference fixed.  In this fixed-\(\gamma_{\rm tr}\), fixed-geometry scan, \(\Xi\) is varied through the microscopic coupling, analogous to controlled parameter rescalings standard in kinetic simulation; \(\Xi\) orders the onset of this family rather than defining full nonlinear similarity~\cite{NykyriPRECompanion}.  The simulation plane is \((x,z)\), with \(x\) along the outflow and \(z\) along the inflow.  Each run uses a Harris sheet~\cite{Harris1962}, a relativistic Boris pusher~\cite{Birdsall1985}, BMT spin precession, the SG force, and a filtered magnetization-current closure
\begin{equation}
\mathbf J_{\rm tot}=\mathbf J_{\rm free}+\mathbf J_M,
\qquad \mathbf J_M=\nabla\times\mathbf M.
\label{eq:Jmclosure}
\end{equation}
The magnetization is coarse-grained before the curl is taken.  This step is essential: the unsmoothed \(\nabla\times\mathbf M\) feedback loop amplifies grid-scale spin noise, whereas a fixed physical coarse-graining length \(\ell\simeq1.1\de\) removes the high-\(k\) numerical branch while preserving the sheet-scale magnetization layer.  The production grid is \(384\times768\), with 200 particles per cell per species, \(\Delta t=0.0025\omega_{pe}^{-1}\), \(L_x\times L_z=25.6\de\times51.2\de\), a weak perturbation magnitude of 0.075, and \(\gamma_{\rm tr}=2\) for every member of the PRL \(\Xi\) scan.  Full implementation details, convergence tests, and energy ledgers are given in the companion PRE paper~\cite{NykyriPRECompanion}; the Supplemental Material summarizes the run table and data products used here~\cite{SM}.

Figure~\ref{fig:xi_sorting} establishes the controlled onset of the spin response.  The classical control contains only weak residual branch imbalance, and \(\Xi=0.1\) remains only weakly organized.  At \(\Xi=0.4\), a coherent two-layer pattern emerges across the sheet; at \(\Xi=0.7\) and 1, the sorting strengthens into multiple sheet-aligned layers.  The common color scale makes the cross-run progression directly comparable, while the scalar measure \(S_\mu\) increases monotonically across the scan.  The transition is therefore not a binary control-versus-active comparison: coherent magnetic-moment sorting turns on progressively with the SG control parameter.

\begin{figure*}[t]
\centering
\safeincludegraphics[width=0.8\textwidth]{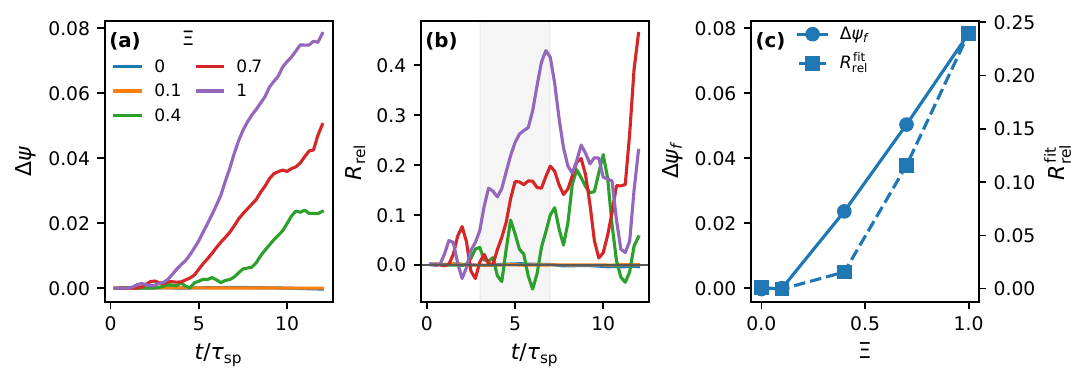}
\caption{\textbf{Relativistically normalized reconnection response across the Stern--Gerlach scan.}
(a) Reconnected-flux increment \(\Delta\psi(t)=\psi(t)-\psi(0)\) for the matched common-\(\gamma_{\rm tr}=2\) weak-seed runs.  (b) Global relativistically normalized flux-growth rate
\(R_{\rm rel}=[d\psi/d(t\omega_{pe})]/(B_{\rm up}V_{A,{\rm rel}})\), obtained from the same dense flux histories using one common smoothing width.  The upstream field and total pair density are measured in symmetric boxes \(|x|\leq6\de\) and \(4\de\leq|z|\leq12\de\); in the cold pair-plasma normalization,
\(V_{A,{\rm rel}}=\sqrt{\sigma/(1+\sigma)}\) with \(\sigma=B_{\rm up}^2/n_{{\rm pair},{\rm up}}\) in code units.  The shaded interval marks the common fit window \(3\leq t/\tausp\leq7\).  (c) Final flux increment \(\Delta\psi_f\) and fitted normalized rate \(R_{\rm rel}^{\rm fit}\) over that interval versus \(\Xi\).  Since \(\psi\) is a global peak-to-peak flux diagnostic, \(R_{\rm rel}\) is a normalized global flux-growth measure rather than a strictly local single-X-line rate when multiple structures are present.}
\label{fig:xi_reconnection}
\end{figure*}

The dynamical response follows the same ordering (Fig.~\ref{fig:xi_reconnection}).  The \(\Xi=0\) and 0.1 runs remain nearly indistinguishable, whereas \(\Xi=0.4\) is the first case to develop sustained flux growth.  Increasing the coupling to \(\Xi=0.7\) and 1 advances the rise of the relativistically normalized flux-growth response.  The instantaneous \(R_{\rm rel}(t)\) curves need not be ordered at every time because the active phases peak at different stages, so the quantitative comparison in Fig.~\ref{fig:xi_reconnection}(c) uses the common-window fit \(R_{\rm rel}^{\rm fit}\) rather than a single smoothed maximum.  Across the active branch, the fitted normalized response and the final flux increment \(\Delta\psi_f\) increase together.  This places the SpinPIC2D response on the same \(B_{\rm up}V_{A,{\rm rel}}\) normalization used in relativistic reconnection studies while preserving the full time-history distinction between onset, growth, and accumulated flux.  Together, Figs.~\ref{fig:xi_sorting} and \ref{fig:xi_reconnection} connect the onset of coherent branch sorting near \(\Xi\simeq0.4\) in the present fixed-\(\gamma_{\rm tr}\) scan with a transition to earlier and faster reconnection.

\begin{figure*}[t]
\centering
\safeincludegraphics[width=0.75\textwidth]{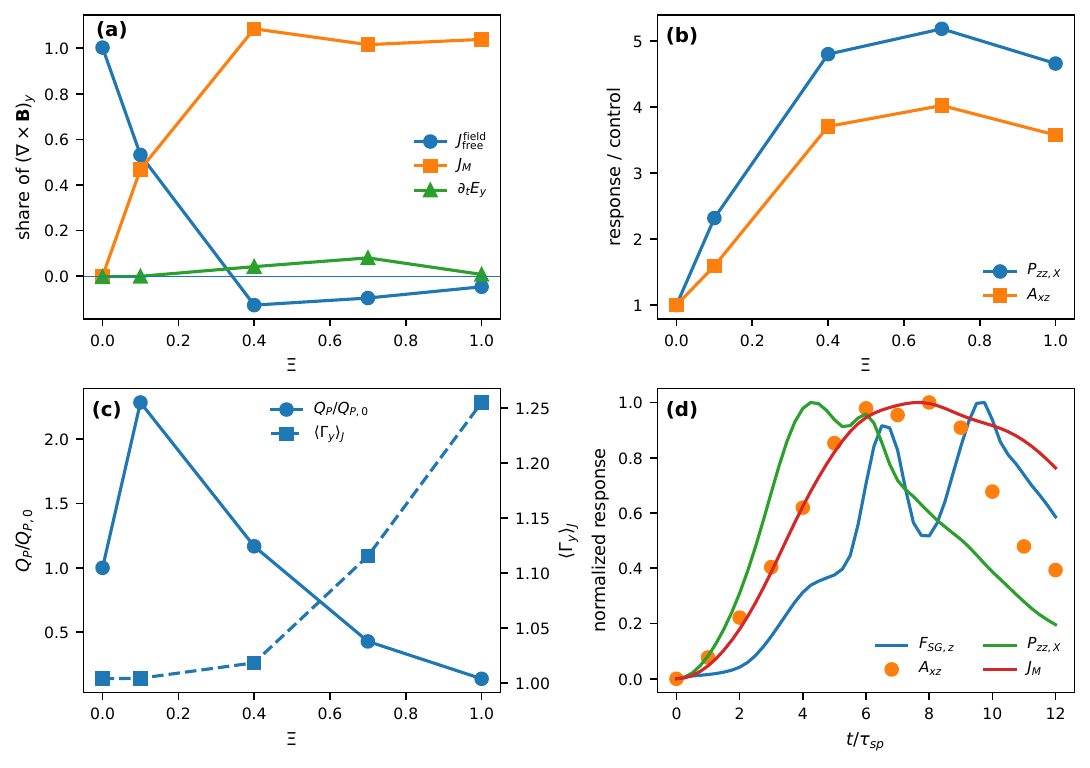}
\caption{\textbf{Spin-mediated kinetic and current-sheet feedback.}
(a) Signed X-line current shares inferred from Maxwell--Amp\`ere balance: the field-balance residual attributed to the free-current channel, \(J_{\rm free}^{\rm field}\), the magnetization current \(J_M\), and the displacement current \(\partial_tE_y\), each normalized to \((\nabla\times\mathbf B)_y\) in the same X-line region.  (b) Sheet-normal pressure proxy \(P_{zz,X}\) (second velocity moment of the distribution) and X-line VDF elongation \(A_{xz}=\sigma_{v_z}/\sigma_{v_x}\), each normalized to the control.  (c) Relative pressure support \(Q_P/Q_{P,0}\), where \(Q_P=P_{zz,X}/(B_{xm}^2/2)\), together with the species-averaged current-weighted carrier Lorentz factor \(\avg{\Gamma_y}_J\).  (d) Normalized temporal responses at \(\Xi=1\) of the sheet-normal SG force, VDF elongation, \(P_{zz,X}\), and \(J_M\).  Panels (a)--(c) are evaluated in the X-line box \(|x|\le1.5\de\), \(|z|\le1.0\de\) at \(t\simeq7\tausp\), species-summed; panel (d) applies the same box at each time.  The ordering supports SG-driven redistribution of velocity space and pressure moments together with a parallel magnetization-current feedback pathway. }
\label{fig:mechanism}
\end{figure*}

The coupled kinetic and current response is summarized in Fig.~\ref{fig:mechanism}.  At \(\Xi=0\), the X-line curl of the magnetic field is supported almost entirely by the field-balance residual attributed to the free-current channel.  At \(\Xi=0.1\), this residual and the magnetization current become comparable, while for \(\Xi\geq0.4\) the inferred balance is magnetization dominated and the residual free-current share becomes weakly counter-directed; the displacement-current contribution remains comparatively small.  This partition is a Maxwell--Amp\`ere reconstruction rather than an independent current-closure test, but it shows how the spin-generated magnetization layer enters the current system.

The current repartition is accompanied by a clear kinetic response.  In Fig.~\ref{fig:mechanism}(b), the increase of the sheet-normal pressure proxy \(P_{zz,X}\), evaluated as the second velocity moment of the distribution, closely tracks the elongation \(A_{xz}=\sigma_{v_z}/\sigma_{v_x}\) of the X-line velocity distribution, with both responses rising rapidly through the same intermediate-\(\Xi\) range.  Yet the pressure increase does not translate into proportionally greater support against the local reconnecting magnetic pressure: Fig.~\ref{fig:mechanism}(c) shows that \(Q_P=P_{zz,X}/(B_{xm}^2/2)\), normalized to the control, falls strongly at large \(\Xi\), while the current-weighted carrier Lorentz factor \(\avg{\Gamma_y}_J\) increases.  This combination provides a bridge to the current-carrier energy and pressure-support constraints identified in relativistic pair reconnection~\cite{GoodbredLiu2022}, while the SG-driven distribution restructuring and magnetization current introduce an additional spin-dependent route to modifying the local balance.

The temporal ordering in Fig.~\ref{fig:mechanism}(d) further argues against a purely serial picture in which \(J_M\) first forms and only then modifies the pressure tensor.  The sheet-normal pressure and VDF elongation respond early, while the magnetization current builds more gradually and remains important during the nonlinear phase; the SG-force diagnostic develops structured peaks over the same interval.  The data therefore support two coupled consequences of the SG force: direct redistribution of the particle distribution and pressure moments, and a parallel \(\mathbf M\rightarrow\mathbf J_M\) electromagnetic feedback pathway.

\begin{figure*}[t]
\centering
\safeincludegraphics[width=0.98\textwidth]{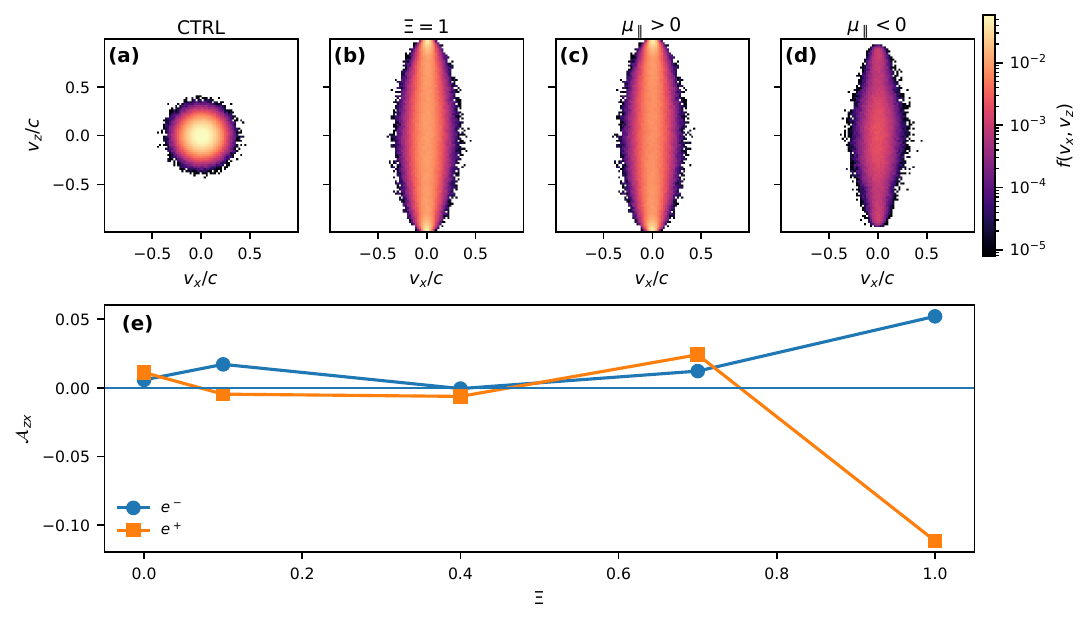}
\caption{\textbf{Spin-resolved velocity-space response and species-resolved \(P_{zx}\) asymmetry.}
(a) Control all-electron \(f(v_x,v_z)\) in the X-line core at \(t\simeq7\tausp\).  (b)--(d) For the \(\Xi=1\) member of the common-\(\gamma_{\rm tr}=2\) scan: all electrons, the \(\mu_\parallel>0\) branch, and the \(\mu_\parallel<0\) branch.  All VDF panels use one logarithmic color scale with limited upper-end saturation, and the branch histograms in (c) and (d) are normalized to the total electron count in the box rather than conditionally within each branch, so the weaker \(\mu_\parallel<0\) panel reflects a smaller branch-integrated population and not merely a narrower conditional distribution.  (e) Electron and positron \(P_{zx}\) asymmetry versus \(\Xi\), quantified by \(\mathcal A_{zx,s}=(\avg{|P_{zx,s}|}_N-\avg{|P_{zx,s}|}_S)/(\avg{|P_{zx,s}|}_N+\avg{|P_{zx,s}|}_S)\) in matched north and south X-line-core regions.  The representative VDFs expose the branch-resolved kinetic deformation directly, while the pressure-tensor scan reveals a species-dependent response and a strong opposite-sign asymmetry at the \(\Xi=1\) endpoint.}
\label{fig:vdf_pzx}
\end{figure*}

A velocity distribution-level view of this response is given in the End Matter: branch-resolved VDFs and a species-dependent \(P_{zx}\) asymmetry (Fig.~\ref{fig:vdf_pzx}) show that the SG force reorganizes velocity space before and during the nonlinear magnetic feedback.  Additional electron generalized-Ohm diagnostics (Fig.~\ref{fig:ohm_jets}) show a temporal handoff in the \(\Xi=1\) case: at \(t\simeq3\tausp\) the structured nonideal field is primarily supported by electron inertia, whereas by \(t\simeq7\tausp\) the pressure-divergence contribution becomes prominent as electron outflow jets and reconnected \(A_y\) topology develop.  The remaining electron-frame residual  \(R_e=E'_y-\mathcal P_e-\mathcal I_e\)  is spatially structured and evolves together with the repartition between free and magnetization currents. Determining whether this residual admits a current-dependent, nonlocal, or wave-mediated closure involving \(\mathbf J_{\rm free}\) or \(\Jm\) is left for future work.

To summarize, the SG force introduces a moment-tagged kinetic anisotropy that is negligible in the solar corona, planetary magnetospheres, and the astrophysical jet environments considered here, can become transitional in some stellar-mass black-hole corona geometries, and becomes important in magnetar-scale current sheets (CS), where near-surface estimates are extrapolative because QED corrections become important.  The controlled scan shows that increasing \(\Xi\) strengthens coherent branch sorting, advances reconnection growth, increases the relativistically normalized flux-growth response, and produces greater accumulated reconnected flux.  The accompanying kinetic diagnostics show that the feedback is not adequately described by a single serial chain.  Because the direct \(y\)-component of the SG force is zero in the 2.5-D geometry, the supported picture is a branched pathway,
\begin{equation*}
F_{\rm SG}\rightarrow \Delta f(v_x,v_z)\rightarrow
\begin{cases}
P_{ij}\ \text{restructuring},\\
\mathbf M\rightarrow\mathbf J_M\rightarrow\text{field/CS feedback},
\end{cases}
\end{equation*}
with the two responses coupled self-consistently through the evolving particle distribution and fields.  The close correspondence between VDF elongation and \(P_{zz}\), the species-dependent \(P_{zx}\) asymmetry, and the transition from free-current to magnetization-dominated Maxwell--Amp\`ere balance provide mutually consistent signatures of that feedback.  This identifies relativistic magnetar current sheets as the clearest natural setting in which spin-kinetic reconnection feedback may be observable.  As a proposed observational consequence rather than a result of these simulations, the near-magnetar regime may exhibit a modification of hard-X-ray polarization through branch-dependent Landau-transition channels~\cite{Sokolov1968,Harding2006}; quantitative Stokes-parameter predictions require radiative-transfer modeling of SG-sorted pair distributions and are deferred to future work.

\section*{End Matter}

\paragraph*{Representative regime survey.}
For the estimates in Table~\ref{tab:survey}, we use \(D_x=10D_z\)~\cite{Heuer2022} and, unless an explicit large-scale geometry is listed, \(D_z=1.5\lambda_{e,{\rm rel}}\)~\cite{Shay1998}, with \(\lambda_{e,{\rm rel}}=\sqrt{\avg{\gamma}}\,c/\omega_{pe}\)~\cite{YuanNarayan2014,Uzdensky2011}.  The transit speed \(\beta_{\rm tr}\) is derived from the electron temperatures or Lorentz factors documented in the Supplemental Material~\cite{SM}.  The table reports \(\Xi_{\rm eff}=P_{\rm eff}\Xi_0\) for \(P_{\rm eff}=0.1\).

\begin{table*}[t]
\caption{Representative values of the SG control parameter.  Geometry: \(\Dx=10\Dz\) and \(\Dz=1.5\lambda_{e,{\rm rel}}\), with \(\lambda_{e,{\rm rel}}=\sqrt{\avg{\gamma}}\,c/\omega_{pe}\), unless an explicit large-scale geometry is listed.  We report \(\XiEff=\Peff\XiZero\) for \(\Peff=0.1\).  In rows using this inertial-scale thickness, the density dependence is \(\XiEff\propto D_x^2/D_z\propto D_z\propto n^{-1/2}\) because \(D_x=10D_z\); hence even large density uncertainties do not move the AGN/blazar/PWN rows near unity.  The near-surface magnetar row is an extrapolative strong-field indicator; QED corrections are expected for fields approaching or exceeding \(B_Q=4.41\times10^9\,\mathrm T\).}
\label{tab:survey}
\begin{ruledtabular}
\begin{tabular}{lcccccc}
Environment & \(\Bin\) (T) & \(n\) (m\(^{-3}\)) & \(\Dz\) (m) & \(\avg{\gamma}\) & \(\btr\) & \(\XiEff(\Peff=0.1)\)\\
\colrule
Earth's magnetotail~\cite{Wang2006,Artemyev2013,Ma2020JA028209} & \(2\times10^{-8}\) & \(10^5\) & \(2.5\times10^{4}\) & \(\approx1\) & \(4.42\times10^{-2}\) & \(\ll1\)\\
Solar corona~\cite{Benz2017,LinForbes2000} & \(10^{-3}\) & \(10^{14}\) & \(8\times10^{-1}\) & \(\approx1\) & \(6.26\times10^{-2}\) & \(\ll1\)\\
Stellar BH corona, large-scale geometry~\cite{YuanNarayan2014,Nattila2024} & \(10^{3}\) & \(10^{22}\) & \(10\) & 3 & 0.94281 & \(\sim0.11\)\\
Stellar BH corona, \(1.5\lambda_{e,{\rm rel}}\)~\cite{YuanNarayan2014,Nattila2024} & \(10^{3}\) & \(10^{22}\) & \(1.4\times10^{-4}\) & 3 & 0.94281 & \(\sim2\times10^{-6}\)\\
M87\(^\ast\) SMBH corona~\cite{EHT2021L13,Kino2014} & \(3\times10^{-3}\) & \(10^{12}\) & \(2.5\times10^{1}\) & 10 & 0.99499 & \(\sim2\times10^{-13}\)\\
Blazar jet, pc scale~\cite{Pushkarev2012,Hovatta2009} & \(4\times10^{-5}\) & \(10^{9}\) & \(9.8\times10^{2}\) & 15 & 0.99778 & \(\ll1\)\\
Blazar jet, inner compact core~\cite{Pushkarev2012,Hovatta2009} & \(9\times10^{-5}\) & \(10^{12}\) & \(3.6\times10^{1}\) & 20 & 0.99875 & \(\ll1\)\\
M87 jet, pc scale~\cite{Reynolds1996,Zamaninasab2014,Mertens2016} & \(2\times10^{-5}\) & \(10^{8}\) & \(2.0\times10^{3}\) & 6 & 0.98601 & \(\ll1\)\\
Magnetar magnetosphere~\cite{Beloborodov2013} & \(10^{7}\) & \(10^{26}\) & \(2.5\times10^{-6}\) & 10 & 0.99499 & \(\sim0.2\)\\
Magnetar surface~\cite{Beloborodov2013,Thompson1993} & \(10^{10}\) & \(10^{28}\) & \(1.1\times10^{-7}\) & 2 & 0.86603 & \(\sim4\times10^{5}\)\\
Crab PWN~\cite{Rees1974,Lyutikov2019,Werner2019} & \(3\times10^{-8}\) & \(10^{8}\) & \(2.5\times10^{5}\) & \(10^5\) & \(\approx1\) & \(\ll1\)\\
\end{tabular}
\end{ruledtabular}
\end{table*}

\paragraph*{Branch-resolved velocity-space response.}
The control X-line-core VDF is compact in the \((v_x,v_z)\) plane, whereas the \(\Xi=1\) distribution is strongly elongated along \(v_z\).  The branch decomposition shows that the \(\mu_\parallel>0\) population carries most of the broad elongated phase-space support, while the \(\mu_\parallel<0\) population is weaker.  The corresponding \(P_{zx}\) asymmetry is species dependent rather than a smooth single-sign scaling with \(\Xi\): the strongest-coupling endpoint develops a pronounced opposite-sign electron--positron response.  The VDF and pressure-tensor diagnostics therefore provide complementary evidence that the SG force reorganizes velocity space before and during the nonlinear magnetic feedback.

\begin{figure*}[t]
\centering
\safeincludegraphics[width=0.98\textwidth]{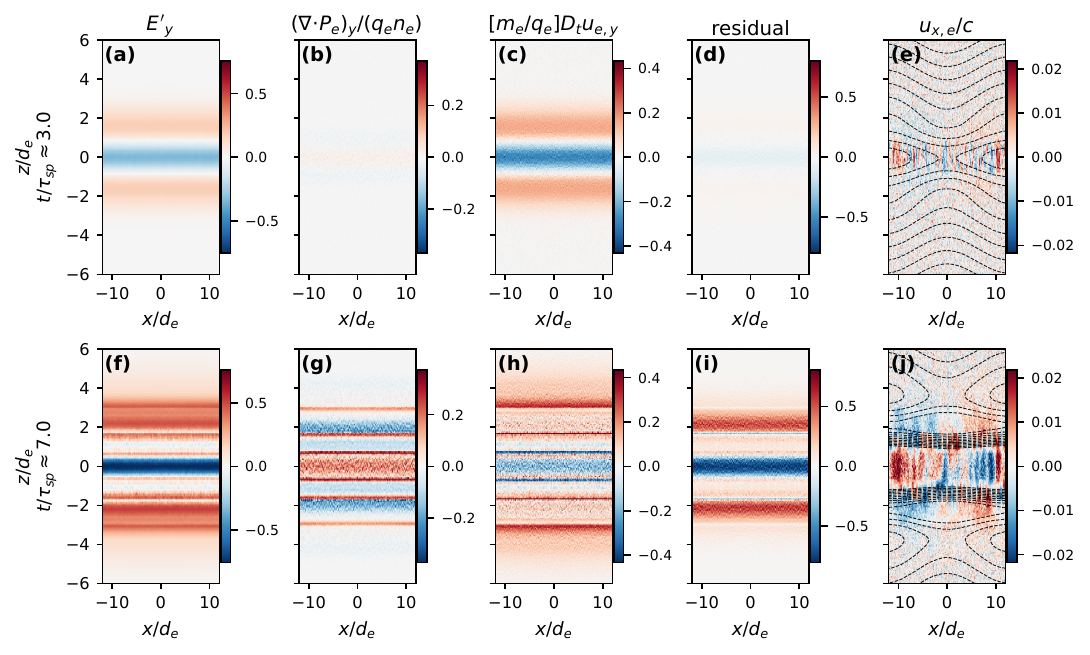}
\caption{\textbf{Representative electron generalized-Ohm terms and outflow jets.}
Top row: \(t\simeq3\tausp\); bottom row: \(t\simeq7\tausp\), for the \(\Xi=1\) member of the common-\(\gamma_{\rm tr}=2\) scan.  Columns show the (a) electron-frame nonideal field \(E'_y=E_y+(\mathbf u_e\times\mathbf B)_y\), (b) the pressure-divergence contribution \(\mathcal P_e=(\nabla\cdot\mathbf P_e)_y/(q_en_e)\) with the electron charge signed (\(q_e=-e\)), (c) the electron inertia contribution \(\mathcal I_e=(m_e/q_e)D_tu_{e,y}\), the (d) residual \(R_e=E'_y-\mathcal P_e-\mathcal I_e\), and (e) the electron bulk flow velocity \(u_{x,e}/c\) with reconstructed \(A_y\) contours.  At the earlier time, the spatial structure of \(E'_y\) is primarily supported by electron inertia; by \(t\simeq7\tausp\), the pressure-divergence contribution becomes prominent (g) and strongly structured while electron outflow jets and reconnected topology are present.  The explicit \(y\)-directed SG term vanishes in the 2.5-D geometry and is therefore not shown.  The residual is retained as an empirical closure diagnostic and motivates projection onto both free and magnetization currents. By $t\simeq 7\tau_{\rm sp}$,  \(u_{x,e}/c\) develops layered, counterstreaming channels (j) associated with the nonlinear, multi-X-line topology. Once reconnection produces appreciable $x$ gradients of $\boldsymbol{\mu}\cdot\mathbf{B}$, the direct
in-plane SG component $F_{{\rm SG},x}=\partial_x(\boldsymbol{\mu}\cdot\mathbf{B})$
may also contribute locally to the oppositely directed electron flows.}
\label{fig:ohm_jets}
\end{figure*}

\paragraph*{Generalized Ohm-law evolution and structured electron flows.}
Figure~\ref{fig:ohm_jets} shows representative electron generalized-Ohm terms for the \(\Xi=1\) member of the common-\(\gamma_{\rm tr}=2\) scan at \(t\simeq3\tausp\) and \(7\tausp\).  The earlier time is primarily supported by structured electron-inertial contributions to \(E'_y\), while the later time output exhibits a stronger pressure-divergence contribution together with developed electron outflows.  Throughout, the electron charge is signed, \(q_e=-e\), and the pressure-divergence term is \(\mathcal P_e=(\nabla\cdot\mathbf P_e)_y/(q_en_e)\), matching the convention of the companion PRE paper~\cite{NykyriPRECompanion}.  The inertia term is evaluated in the bulk-velocity form \(\mathcal I_e=(m_e/q_e)D_tu_{e,y}\); a fully relativistic, enthalpy-weighted momentum balance is left for future work, so the decomposition is presented as a structural diagnostic rather than an exact closure.  Because \(F_{{\rm SG},y}=0\) identically in the 2.5-D geometry, the direct SG term is not plotted; the SG channel enters indirectly through the \(x\)-\(z\) redistribution and the self-consistent magnetization-current feedback.  We define the plotted remainder as \(R_e=E'_y-\mathcal P_e-\mathcal I_e\) and label it simply \(R_e\); no closure fit is performed here.  Its structured sheet-scale morphology motivates testing, in future work, the empirical closure \(R_e\simeq\eta_1J_{y,{\rm free}}+\eta_2J_{M,y}\); such coefficients would quantify projection onto the free- and magnetization-current channels, e.g., via current-dependent resistivity models~\cite{NykyriOtto2001} rooted in wave--particle interactions.  The impact of the SG/magnetization-current channel on plasma wave modes relevant to reconnection, or the excitation of new modes, is likewise deferred.

\clearpage
\onecolumngrid
\begin{center}
\textbf{Supplemental Material for: Stern--Gerlach Spin Sorting and Dynamical Feedback in Relativistic Pair-Plasma Reconnection\\}
  K. Nykyri
  \end{center}

This Supplemental Material gives the branch-resolved SG derivation, the pair-plasma sign map, the transit-speed closures used in Table~I of the End Matter, Harris-sheet equilibrium verification for the complete five-point production scan of $\Xi$, and the production-run/data-product summary for the energy-gated SpinPIC2D simulations using the complete SG-force-plus-magnetization-current closure.  All members of the controlled PRL scan share the common transit/coupling anchor $\gamma_{\rm tr}=2$.  This Supplemental Material retains only verification and provenance products that directly support the Letter.  Figure~S1 verifies the common underlying Harris equilibrium and common seeded X-line geometry across the full five-point $\Xi$ scan.  The production-run table documents the exact data sets used by Figs.~2--5 and End-Matter Fig.~6.  Detailed algorithmic validation, energy-convergence matrices, high-$k$ magnetization-feedback diagnosis, smoothing-length scans, time-step refinement, and relativistic-population sweeps are reported in the companion PRE paper~\cite{NykyriPRECompanion} rather than duplicated here.

\section{Branch-resolved Stern--Gerlach displacement}

For electrons,
\begin{equation}
\bm\mu_e=-\frac{g_s\mu_B}{\hbar}\mathbf S,
\qquad g_s\simeq2,
\qquad \mu_B=\frac{e\hbar}{2m_e}.
\label{eq:sg_mu}
\end{equation}
In the local plasma frame, the SG force follows from the dipole interaction,
\begin{equation}
\mathbf F_{\rm SG}=\nabla(\bm\mu\cdot\mathbf B).
\label{eq:sg_force}
\end{equation}
For a 2.5-D reconnection geometry with guide-field direction \(\hat y\), write
\(\mathbf B=b_x\hat x+b_y\hat y+b_z\hat z\).  The electron sheet-normal component is then
\begin{equation}
F_{S,z}=-\frac{g_s\mu_B}{\hbar}\left(
S_x\frac{\partial b_x}{\partial z}
+S_y\frac{\partial b_y}{\partial z}
+S_z\frac{\partial b_z}{\partial z}\right),
\label{eq:sg_fzfull}
\end{equation}
with analogous expressions for \(F_{S,x}\) and \(F_{S,y}\).

In the present scaling estimate the dominant inflow field \(\Bin\hat x\) sets the leading BMT precession axis.  In the local instantaneous rest frame the spin obeys
\begin{equation}
\frac{d\mathbf S}{d\tau}=\bm\Omega_{\rm BMT}\times\mathbf S,
\label{eq:sg_bmt}
\end{equation}
where we retain only the dominant magnetic contribution to \(\bm\Omega_{\rm BMT}\).  The proper-time precession frequency is therefore
\begin{equation}
\Omega_{\rm BMT}^{(\tau)}\simeq \frac{g_s\mu_B\Bin}{\hbar}.
\label{eq:sg_omega_tau}
\end{equation}
Because the current-sheet residence time \(\Tr\) is measured in the reconnection frame, the relevant comparison uses the coordinate-time frequency
\begin{equation}
\Omega_{\rm BMT}^{(t)}=\frac{\Omega_{\rm BMT}^{(\tau)}}{\gamma}
\simeq \frac{g_s\mu_B\Bin}{\gamma\hbar}.
\label{eq:sg_omega_t}
\end{equation}
This is not intended as a full covariant spin-transport solution; it is the minimal local-frame estimate needed to justify precession averaging when \(\Omega_{\rm BMT}^{(t)}\Tr\gg1\).

Let \(\theta\) be the polar angle between \(\mathbf S\) and \(\hat x\).  During precession about \(\hat x\), the spin components may be written as
\begin{align}
\avg{S_x} &= \frac{\hbar}{2}\cos\theta,
\label{eq:sg_sx}\\
\avg{S_y} &= \frac{\hbar}{2}\sin\theta\cos(\Omega_{\rm BMT}^{(t)}t),
\label{eq:sg_sy}\\
\avg{S_z} &= -\frac{\hbar}{2}\sin\theta\sin(\Omega_{\rm BMT}^{(t)}t).
\label{eq:sg_sz}
\end{align}
Averaging over one precession period \(T_\Omega=2\pi/\Omega_{\rm BMT}^{(t)}\) gives
\begin{align}
\avg{S_y}_{T_\Omega}
&=\frac{1}{T_\Omega}\int_0^{T_\Omega}
\frac{\hbar}{2}\sin\theta\cos(\Omega_{\rm BMT}^{(t)}t)\,dt=0,
\label{eq:sg_syavg}\\
\avg{S_z}_{T_\Omega}
&=-\frac{1}{T_\Omega}\int_0^{T_\Omega}
\frac{\hbar}{2}\sin\theta\sin(\Omega_{\rm BMT}^{(t)}t)\,dt=0.
\label{eq:sg_szavg}
\end{align}
Thus only the spin projection onto the precession axis survives the EDR transit average.  Substituting Eqs.~(\ref{eq:sg_syavg})--(\ref{eq:sg_szavg}) into Eq.~(\ref{eq:sg_fzfull}) yields
\begin{equation}
\avg{F_{S,z}}
=-\frac{g_s\mu_B}{\hbar}\avg{S_x}\frac{\partial b_x}{\partial z}.
\label{eq:sg_fzavg}
\end{equation}
The corresponding averaged in-plane and guide-direction components are
\begin{align}
\avg{F_{S,x}}&=-\frac{g_s\mu_B}{\hbar}\avg{S_x}\frac{\partial b_x}{\partial x},
\label{eq:sg_fxavg}\\
\avg{F_{S,y}}&=-\frac{g_s\mu_B}{\hbar}\avg{S_x}\frac{\partial b_x}{\partial y}.
\label{eq:sg_fyavg}
\end{align}

For the sign convention used in Fig.~1 of the main text, the reconnecting field near the sheet center is
\begin{equation}
B_x\simeq +\Bin\frac{z}{\Dz},
\qquad
\frac{\partial B_x}{\partial z}=+\frac{\Bin}{\Dz}>0.
\label{eq:sg_harris_gradient}
\end{equation}
Equation~(\ref{eq:sg_fzavg}) then gives
\begin{equation}
\avg{F_{S,z}}
=-\frac{g_s\mu_B\Bin}{\hbar\Dz}\avg{S_x}.
\label{eq:sg_fz_harris}
\end{equation}
For electrons, \(\avg{S_x}=+\hbar/2\) gives \(\mu_\parallel<0\) and a downward force, while \(\avg{S_x}=-\hbar/2\) gives \(\mu_\parallel>0\) and an upward force.  Expressed in terms of the field-aligned magnetic-moment projection \(\mu_\parallel=\bm\mu\cdot\hat{\mathbf B}_{\rm in}\), the convention-independent branch force is
\begin{equation}
F_{{\rm SG},z}\simeq \mu_\parallel\frac{\Bin}{\Dz}
=\operatorname{sgn}(\mu_\parallel)\eta_\mu\mu_B\frac{\Bin}{\Dz},
\qquad
\eta_\mu\equiv\frac{|\mu_\parallel|}{\mu_B}.
\label{eq:sg_branch_force}
\end{equation}
The magnitude estimate used in the main text is recovered by replacing the branch factor \(\eta_\mu\) with the effective participating/ensemble factor \(\Peff\).

Using the constant-$\gamma$ relativistic-inertia estimate $a_z=F_{{\rm SG},z}/(\avg{\gamma}m_e)$ over \(\Tr=\Dx/(2\btr c)\) gives the signed branch displacement
\begin{equation}
\avg{z}_{\mu_\parallel}=
\operatorname{sgn}(\mu_\parallel)\eta_\mu
\frac{\mu_B\Bin\Dx^2}{8\avg{\gamma}m_e\Dz\btr^2c^2}.
\label{eq:sg_branch_z}
\end{equation}

The effective participation factor $P_{\rm eff}$
generalizes the single-particle branch factor $\eta_\mu$
to an ensemble.  Operationally,
$P_{\rm eff}=\langle C\eta_\mu\rangle_n$, where
$C\in[0,1]$ represents coherent transit participation and
$\eta_\mu=|\mu_{\parallel}|/\mu_B$ represents the surviving
magnetic-moment projection.  For a fully projected,
fully participating population, $C=\eta_\mu=1$ and
$P_{\rm eff}=1$.
For a thermally broadened spin distribution in an
unpolarized plasma the net ($\mu_\parallel$-averaged)
center-of-mass force vanishes; the branch-resolved
centroids of the $\mu_\parallel>0$ and $\mu_\parallel<0$
sub-populations, however, separate according to Eq.~(\ref{eq:sg_branch_z})
even when $\langle\mu_\parallel\rangle=0$, because the
averaging is performed within each branch separately.
This distinction—between a net macroscopic observable,
which requires non-zero ensemble polarization, and
branch-resolved sorting, which does not—is the central
physical point of the diagnostic design.

Finite gradients along \(x\) or \(y\) decorate this primary sheet-normal sorting:
\begin{align}
\avg{x}_{\mu_\parallel}&\sim
\operatorname{sgn}(\mu_\parallel)\eta_\mu
\frac{\mu_B}{2\avg{\gamma}m_e}
\frac{\partial b_x}{\partial x}\Tr^2,
\label{eq:sg_branch_x}\\
\avg{y}_{\mu_\parallel}&\sim
\operatorname{sgn}(\mu_\parallel)\eta_\mu
\frac{\mu_B}{2\avg{\gamma}m_e}
\frac{\partial b_x}{\partial y}\Tr^2.
\label{eq:sg_branch_y}
\end{align}

The in-plane SG displacements in Eqs.~(\ref{eq:sg_branch_x})--(\ref{eq:sg_branch_y}) are retained for analytic completeness, but they are negligible in the Harris-sheet inflow region for two independent reasons.  First, the guide-direction contribution vanishes identically in the 2.5-D simulation plane because \(\partial/\partial y\equiv0\).  Second, the ratio of the in-plane force to the sheet-normal force in the inflow region scales as
\begin{equation}
  \frac{|\avg{F_{S,x}}|}{|\avg{F_{S,z}}|}
  \sim
  \frac{|\partial B_x/\partial x|}{|\partial B_x/\partial z|}
  \sim
  |\epsilon|\frac{\lambda}{L_x}
  \approx 0.075\times\frac{1}{25.6}
  \approx 2.9\times10^{-3},
  \label{eq:sg_inplane_ratio}
\end{equation}
where \(|\epsilon|=0.075\) is the magnitude of the common production flux perturbation, \(\lambda=1d_e\) is the Harris half-width, and \(L_x=25.6d_e\) is the box length.  The in-plane SG force is therefore suppressed by roughly \(10^{-3}\)--\(10^{-2}\) relative to the sheet-normal component throughout the inflow diagnostic region.  Any residual \(x\)-deflection also accumulates along the periodic outflow direction and does not produce a net north--south signal in the branch-sorting diagnostics \(A_\mu\) and \(S_\mu\).  The SpinPIC2D SG pusher computes the nonzero simulation-plane gradient \(\nabla_{xz}(\bm{s}\cdot\bm{B})\) and applies both sheet-normal and in-plane components; the dominance of the \(z\) term is therefore a physical property of the Harris inflow geometry, not a code approximation.

The robust sheet-normal sign statement for \(\partial B_x/\partial z>0\) is
\begin{equation}
\mu_\parallel>0\rightarrow z>0,
\qquad
\mu_\parallel<0\rightarrow z<0.
\label{eq:sg_sign}
\end{equation}

\section{Electron--ion versus pair plasmas}

In an electron--ion plasma, the SG response is dominated by the electron channel because the ion magnetic-moment-to-inertia ratio is much smaller.  In an electron--positron pair plasma, both species participate kinematically.  Let \(s_\parallel=\hat s\cdot\hat{\mathbf B}_{\rm in}=\pm1\), where \(\hat s=2\mathbf S/\hbar\).  Then
\begin{equation}
\mu_{\parallel,e^-}=-\mu_Bs_\parallel,
\qquad
\mu_{\parallel,e^+}=+\mu_Bs_\parallel.
\label{eq:sm8}
\end{equation}
With \(\partial B_x/\partial z>0\),
\begin{align}
&e^-: &&s_\parallel=+1\Rightarrow\mu_\parallel<0\Rightarrow z<0,\nonumber\\
&e^-: &&s_\parallel=-1\Rightarrow\mu_\parallel>0\Rightarrow z>0,\nonumber\\
&e^+: &&s_\parallel=+1\Rightarrow\mu_\parallel>0\Rightarrow z>0,\nonumber\\
&e^+: &&s_\parallel=-1\Rightarrow\mu_\parallel<0\Rightarrow z<0.
\label{eq:sm9}
\end{align}
The sorting is therefore moment-tagged and species-mixed.

\section{Transit-speed closures used in Table I}

For relativistic source classes we use
\begin{equation}
\beta_\gamma=\sqrt{1-\avg{\gamma}^{-2}}.
\label{eq:sm10}
\end{equation}
For nonrelativistic environments we use \(\beta_{\rm th}=\sqrt{k_BT_e/(m_ec^2)}\).  The adopted values are listed in Table~\ref{tab:betas}.

\begin{table}[h]
\caption{Transit-speed values used in Table~I of the main text.}
\label{tab:betas}
\begin{ruledtabular}
\begin{tabular}{lcc}
Environment & closure & \(\btr\)\\
\colrule
Earth magnetotail & \(T_e=1\,\mathrm{keV}\) & \(4.42\times10^{-2}\)\\
Solar corona & \(T_e=2\,\mathrm{keV}\) & \(6.26\times10^{-2}\)\\
Stellar BH corona & \(\beta_\gamma(\avg\gamma=3)\) & 0.94281\\
M87\(^\ast\) SMBH corona & \(\beta_\gamma(\avg\gamma=10)\) & 0.99499\\
Blazar jet, pc scale & \(\beta_\gamma(\avg\gamma=15)\) & 0.99778\\
Blazar jet, inner core & \(\beta_\gamma(\avg\gamma=20)\) & 0.99875\\
M87 jet & \(\beta_\gamma(\avg\gamma=6)\) & 0.98601\\
Magnetar magnetosphere & \(\beta_\gamma(\avg\gamma=10)\) & 0.99499\\
Magnetar surface & \(\beta_\gamma(\avg\gamma=2)\) & 0.86603\\
Crab PWN & \(\beta_\gamma(\avg\gamma=10^5)\) & \(\approx1\)\\
\end{tabular}
\end{ruledtabular}
\end{table}

With \(\Dx=10\Dz\) and \(\Dz=1.5\lambda_{e,{\rm rel}}\), \(\lambda_{e,{\rm rel}}=\sqrt{\avg{\gamma}}\,c/\omega_{pe}\), \(\XiEff\propto\Dz\propto n^{-1/2}\) at fixed \(\avg{\gamma}\).  This scaling is used when assessing density uncertainty in the jet rows.

\section{BMT coherence check}

The number of BMT precession cycles during one current-sheet transit is
\begin{equation}
N_{\rm prec}=\frac{\Omega_{\rm BMT}^{(t)}\Tr}{2\pi}
\simeq \frac{g_s\mu_B\Bin}{2\pi\avg\gamma\hbar}\Tr.
\label{eq:sm11}
\end{equation}

This estimate neglects guide-field gradients
($\partial B_y/\partial z$), which contribute subdominantly
to $\bm{\Omega}_{\rm BMT}$ in the Harris geometry, and uses
a constant-$\gamma$, straight-transit closure for the trajectory.
It is intended as an order-of-magnitude coherence test, not
as a covariant orbit solution; no assumption is made that the
trajectory corrections are perturbatively small for
$\beta_{\rm tr}\simeq1$.
For the environments considered here the magnetic precession is fast on the transit time at the order-of-magnitude level, justifying the averaging of transverse spin components.  A complete spin-transport calculation is beyond the scope of this Letter.

\subsection*{Density and field-strength uncertainties}

Entries in Table~I of the main text for black-hole
coronae, AGN jets, and the magnetar magnetosphere span
environments where observational constraints on $n$ and
$B_{\rm in}$ carry uncertainties of one to two orders of
magnitude.
With the inertial-scale geometry $D_z=1.5\lambda_{e,\rm rel}$
and $\lambda_{e,\rm rel}\propto n^{-1/2}$, the scaling
$\Xi_{\rm eff}\propto D_z\propto n^{-1/2}$ holds at fixed
$\langle\gamma\rangle$.
A factor-of-100 density uncertainty therefore shifts
$\Xi_{\rm eff}$ by only a factor of 10.
For the AGN/blazar and PWN rows, which already yield
$\Xi_{\rm eff}\ll1$ by three to nine orders of magnitude,
no plausible density revision brings them near the
transitional regime.
For the magnetar rows, a factor-of-100 density increase
reduces $\Xi_{\rm eff}$ by one decade, leaving the outer
magnetosphere sub-transitional but the surface row still at
$\Xi_{\rm eff}\sim10^4$.
The near-surface row should be read as an extreme
extrapolation: pair-cascade rates and photon splitting
become important for $B\gtrsim B_Q=4.41\times10^9$~T,
and a self-consistent treatment is beyond the scope of this
Letter.

\subsection*{Choice and definition of $P_{\rm eff}$}

The effective participation factor $P_{\rm eff}\in(0,1]$
combines coherent transit participation and the surviving
magnetic-moment projection.  Operationally,
$P_{\rm eff}=\langle C\eta_\mu\rangle_n$, where
$C\in[0,1]$ represents coherent participation and
$\eta_\mu=|\mu_\parallel|/\mu_B$.
In a fully polarized, fully participating beam,
$C=\eta_\mu=P_{\rm eff}=1$.
In an unpolarized ensemble with an isotropic spin distribution,
the net center-of-mass force vanishes, but the branch-resolved
centroids separate whenever $\Xi_0>0$, because each
$\mu_{\parallel}$ branch individually satisfies
Eq.~(\ref{eq:sg_branch_z}) regardless of the ensemble polarization.
For the representative regime survey, we adopt
$P_{\rm eff}=0.1$ as a fiducial benchmark; the fully projected
value $P_{\rm eff}=1$ raises every entry in Table~I of the main
text by one decade.
The critical effective branch factor
$P_{{\rm eff},c}=\Xi_0^{-1}$ gives the minimum effective
participation/projection factor required for $\Xi_{\rm eff}=1$;
for the magnetar surface row,
$P_{{\rm eff},c}\sim2\times10^{-6}$, meaning SG
transport is strong even if the vast majority of moments
are randomized.

\section{SpinPIC2D model}

This Letter uses the complete implementation of SpinPIC2D with SG force and magnetization-current feedback.  The model is a 2.5-dimensional relativistic pair-plasma PIC code in the $(x,z)$ plane.  Particles are advanced in momentum space with a relativistic Boris pusher~\cite{Birdsall1985}; spins are advanced by a BMT rotation; the SG force is applied as a momentum kick proportional to the simulation-plane gradient $\nabla_{xz}(\mathbf s\cdot\mathbf B)$; and the electromagnetic fields are advanced on a Yee grid.  The common PRL scan holds $\gamma_{\rm tr}=2$ fixed while varying $\Xi$.

The complete SG-force-plus-magnetization-current production branch, denoted FullM7 in the campaign metadata, uses the current entering Amp\`ere's law
\begin{equation}
\mathbf J_{\rm tot}=\mathbf J_{\rm free}+\mathbf J_M,
\qquad \mathbf J_M=\nabla\times\mathbf M,
\label{eq:si_jm}
\end{equation}
where $\mathbf M$ is computed from local spin-weighted particle moments and coarse-grained at a fixed physical scale before the curl is evaluated.  The magnetization curl is evaluated with the same energy-adjoint Yee-grid convention used by the field update.  The coarse-graining length is fixed for a production run and is not a time-dependent dissipation parameter: it suppresses the nonphysical grid-scale SG--magnetization feedback branch identified in the convergence study while preserving the sheet-scale $J_{M,y}$ layer.  The complete derivation, update ordering, filter-length scan, grid/PPC/time-step tests, and energy ledgers are reported in the companion PRE paper~\cite{NykyriPRECompanion}.

For clarity, three current diagnostics appearing in the production outputs should not be conflated.  The HDF5 quantity \texttt{/current/Jy\_free} is the legacy particle-deposited free current; as stored, it is an integrated-cell moment and is converted to a density-like current by division by $\Delta A=\Delta x\Delta z$.  The production field update can use the work-conjugate current, whose exact spatial field was not stored in the existing HDF5 snapshots.  Therefore the current share shown in Fig.~4(a) of the Letter uses the Maxwell--Amp\`ere reconstruction: 
\begin{equation}
J_{\rm free}^{\rm field}
=(\nabla\times\mathbf B)_y-J_{M,y}-\partial_tE_y,
\label{eq:si_jfree_field}
\end{equation}

\subsection{Harris-equilibrium and common-initialization verification}

\begin{figure}[h]
\centering
\safeincludegraphics[width=0.98\textwidth]{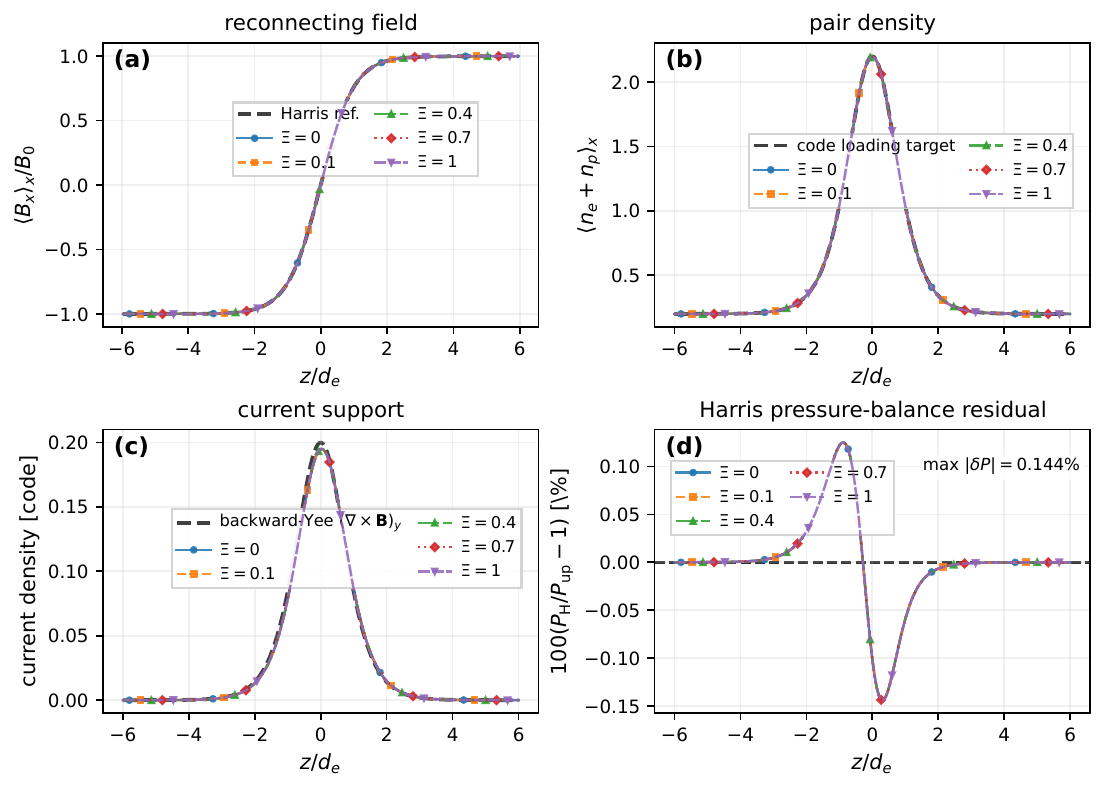}
\caption{\textbf{Harris-equilibrium and common-initialization verification for the matched PRL production scan.}
Initial profiles for the five common-\(\gamma_{\rm tr}=2\) cases, \(\Xi=0,0.1,0.4,0.7,1\).  (a) The \(x\)-averaged reconnecting field from the common seeded \(t=0\) state compared with the unperturbed Harris reference \(B_{\rm H}/B_0=\tanh(z/\lambda)\); the imposed flux perturbation averages away in \(x\) in this profile.  (b) Deposited total pair density compared with the SpinPIC2D loading target \(2[\mathrm{sech}^2(z/\lambda)+n_{\rm bg}]\).  (c) Deposited legacy free-current density compared with the exact backward-Yee magnetic curl.  This is an initialization/profile check rather than an independent work-current closure test.  (d) Pressure-balance residual evaluated with the same thermodynamic prescription as the code initialization and equilibrium audit, \(P_{\rm H}(z)=n_{\rm pair}(z)T+B_{\rm H}^2(z)/2\), with \(T=\beta/2\).  The residual is of order \(10^{-3}\) in fractional pressure.  All production cases additionally contain the same finite flux perturbation, \(\epsilon=-0.075\) in the code convention, which creates the central X-line saddle geometry; the fully seeded two-dimensional \(t=0\) field is therefore not expected to satisfy the unperturbed one-dimensional Harris pressure-balance relation pointwise.  The overlap of the five scan members verifies a common macroscopic Harris loading and common seeded initial geometry across the controlled \(\Xi\) scan.}
\label{fig:si_harris}
\end{figure}

The controlled $\Xi$ scan changes the spin coupling while holding the macroscopic particle loading and seeded magnetic geometry fixed.  The underlying one-dimensional pair-Harris equilibrium is initialized with
\begin{equation}
T=\frac{\beta}{2},\qquad
P_{\rm H}(z)=n_{\rm pair}(z)T+\frac{B_{\rm H}^2(z)}{2},
\qquad
B_{\rm H}(z)=B_0\tanh\!\left(\frac{z}{\lambda}\right),
\label{eq:si_harris_pressure}
\end{equation}
where the pair loading satisfies \(2n_0T=B_0^2/2\).  The production cases then apply the same finite magnetic-flux perturbation, \(\epsilon=-0.075\) in the code convention, to establish the central X-line saddle geometry.  Consequently, the fully seeded two-dimensional \(t=0\) magnetic field is not expected to satisfy the unperturbed one-dimensional Harris pressure-balance relation pointwise.  Figure~\ref{fig:si_harris} therefore evaluates the equilibrium pressure residual using the same unperturbed Harris field \(B_{\rm H}\) and thermodynamic prescription \(n_{\rm pair}T\) used by the SpinPIC2D initialization and initial-equilibrium audit, while the field, density, and current panels verify the common seeded production state directly.

\subsection{Run set used in the Letter figures}

The weak-seed $\Xi$ scan used in Figs.~2--5 and the End-Matter Fig.~6 consists of the following runs.  Every member uses the common transit/coupling anchor $\gamma_{\rm tr}=2$.  In this scan, $\gamma_{\rm tr}=2$ serves as the transit parameter entering the definition of $\Xi$ while the particle loading is the common Harris thermal loading (measured mean initial Lorentz factor $\langle\gamma\rangle_0\approx1.015$); it is therefore distinct from the
explicitly loaded $\langle\gamma\rangle_0\simeq2.01$ relativistic-sweep case of the companion PRE paper~\cite{NykyriPRECompanion}, which shares the same $\Xi=1$ and $\chi_{\rm sim}$ but a different particle distribution. The $\Xi=0$ run is the classical relativistic PIC control; the $\Xi>0$ cases use the full SG-force-magnetization-current closure (FullM7).  The $\Xi=1$ endpoint is the run named A4 in the campaign directory deposited in Figshare \url{https://doi.org/10.6084/m9.figshare.32983562}.

\begin{table}[h]
\caption{Production runs used in the Letter and End Matter.  All runs use the common $\gamma_{\rm tr}=2$ anchor, a $384\times768$ grid, 200 particles per cell per species, $L_x\times L_z=25.6d_e\times51.2d_e$, $\Delta t=0.0025\omega_{pe}^{-1}$, the common weak flux perturbation $\epsilon=-0.075$ in the code convention (central X-line saddle), and the same underlying Harris-sheet loading. The FullM7 runs use $\ell=1.10d_e$ magnetization coarse graining.}
\label{tab:si_prl_runs}
\begin{ruledtabular}
\begin{tabular}{lccc}
Case & directory label & physics & role\\
\colrule
\(\Xi=0\) & PRL\_A3\_CTRL\_pertm0p075\_tau12 & classical relativistic PIC & control\\
\(\Xi=0.1\) & PRL\_XI\_0p1\_FULL\_ell1p10\_pertm0p075\_tau12 & FullM7 & subthreshold spin run\\
\(\Xi=0.4\) & PRL\_XI\_0p4\_FULL\_ell1p10\_pertm0p075\_tau12 & FullM7 & onset/intermediate run\\
\(\Xi=0.7\) & PRL\_XI\_0p7\_FULL\_ell1p10\_pertm0p075\_tau12 & FullM7 & strong-coupling run\\
\(\Xi=1.0\) & PRL\_A4\_FULL\_Xi1\_ell1p10\_pertm0p075\_tau12 & FullM7 & endpoint run\\
\end{tabular}
\end{ruledtabular}
\end{table}

\subsection{Energy gate and validation status}

All production points used in the Letter remain below the adopted one-percent total-energy-drift gate.  The control run has negligible drift; the \(\Xi=1\) endpoint has the largest drift, approximately \(0.37\%\), while still remaining well inside the validation gate.  Strong-seed support runs with \(\ell=1.10d_e\) and \(\ell=0.99d_e\) show the same qualitative enhancement and provide filter-sensitivity checks.  A long no-spin control run demonstrates that the underlying relativistic PIC/Yee solver remains stable over much longer integration times; it is a code-stability check and is not used as a primary physics point in this Letter.

The final validation separates three issues: (i) the classical relativistic solver exhibits negligible operational energy drift; (ii) SG-force-only diagnostic tests can isolate branch sorting but are not used as production physics points because they omit the reciprocal magnetization-current channel; and (iii) the complete SG+$\nabla\times\mathbf M$ closure requires fixed-length magnetization coarse graining to remove the nonphysical high-$k$ feedback branch.  The PRE companion gives the full convergence sequence and explains why the final PRL production corner uses \(384\times768\), PPC200, \(\Delta t=0.0025\), and \(\ell=1.10d_e\).

\subsection{Ohm-law, current-partition, and VDF products}

Native electron Ohm-law products are written to the production snapshots as $E'_y$, pressure-divergence, inertia, direct SG, and residual terms.  With $\partial/\partial y=0$ in the production geometry, the direct $y$ component of the SG force vanishes identically even when SG physics is active.  The spin channel therefore enters the $y$-directed reconnection system indirectly through sheet-normal velocity-space redistribution and the self-consistent magnetization current $J_{M,y}$.  The End-Matter Fig.~6 uses the $\Xi=1$ snapshots at
$t\simeq3\tau_{\rm sp}$ and $7\tau_{\rm sp}$ to show the evolution
from early electron-inertial support toward a stronger pressure-divergence contribution, together with a nonlinear, layered electron bulk-flow structure and reconstructed $A_y$ topology.  The plotted remainder is
\begin{equation}
R_e=E'_y-\mathcal P_e-\mathcal I_e,
\label{eq:si_ohm_residual}
\end{equation}
where $\mathcal P_e=(\nabla\cdot\mathbf P_e)_y/(q_en_e)$ with $q_e=-e$, and $\mathcal I_e=(m_e/q_e)D_tu_{e,y}$.  The residual remains spatially structured.  Determining whether it admits a current-dependent, nonlocal, or wave-mediated closure involving $J_{\rm free}$ and $J_M$ is left for future work rather than assumed in the present Letter.

The current-partition diagnostic underlying Fig.~4(a) uses Eq.~(\ref{eq:si_jfree_field}) and the exact backward-Yee curl.  Consequently, $J_{\rm free}^{\rm field}$ is a field-balance reconstruction.  The deposited free current is retained in the HDF5 files for provenance and auxiliary comparisons, but it is not substituted for the field-inferred quantity in the published partition panel.

The region-resolved VDF products saved in the production snapshots include, for electrons and positrons in the X-line core, left/right exhausts, and north/south upstream regions,
\begin{equation*}
\mathrm{f}_{v_xv_z},\quad
\mathrm{f}_{v_xv_y},\quad
\mathrm{f}_{v_\parallel v_\perp},\quad
\mathrm{f}_{v_xv_z}^{\mu_\parallel>0},\quad
\mathrm{f}_{v_xv_z}^{\mu_\parallel<0}.
\end{equation*}
The Letter uses the X-line-core $v_x$--$v_z$ products and their $\mu_\parallel$ branch decomposition at $t\simeq7\tau_{sp}$.  More localized $v_y$--$v_z$ and branch-resolved $v_x$--$v_y$ diagnostics are natural follow-up products for a dedicated plasma-parameter and wave-particle-interaction study.

\section{Data and code availability}

The simulation outputs, exact HDF5 snapshots, derived analysis tables, figure-generation scripts, run manifests, and checksums supporting this Letter are available in the versioned Figshare repository:
\url{https://doi.org/10.6084/m9.figshare.32983562}

\end{document}